%% file: all-categorical-symmetries.tex
\documentclass[a4paper,11pt]{article}
\pdfoutput=1
\usepackage{jheppub}
\usepackage{braket, tikz, tikz-cd, caption}
\usepackage{indentfirst}
\input{preamble}

\definecolor{darkred}{rgb}{0.5,0.15,0.15}
\hypersetup{colorlinks=true,urlcolor=darkred,linkcolor=darkred,citecolor=darkred}

\title{Towards All Categorical Symmetries in 2+1 Dimensions}
\author{Mathew Bullimore,}
\author{Jamie J. Pearson}
\affiliation{Department of Mathematical Sciences, Durham University, \\
Upper Mountjoy, Stockton Road, Durham, DH1 3LE, United Kingdom}
\emailAdd{mathew.r.bullimore@durham.ac.uk}
\emailAdd{jamie.j.pearson@durham.ac.uk}

\abstract{We investigate the most general gauging operations in 2+1 dimensional oriented field theories with finite symmetry groups, which correspond to gapped boundary conditions in 3+1 dimensional Dijkgraaf-Witten theory. The classification is achieved by enumerating 2+1 dimensional oriented topological quantum field theories that cancel the 't Hooft anomaly associated with the symmetry. This framework is rigorously formulated using twisted crossed extensions of modular fusion categories and projective 3-representations. Additionally, we explore the resulting fusion 2-category symmetries and argue that this framework captures all possible categorical symmetries in 2+1 dimensional oriented field theories.}

\begin{document}
\maketitle



\section{Introduction}
\label{sec:intro}


\subsection{Motivation}
\label{subsec:motivation}

The method of gauging finite invertible symmetries has played a pivotal role in the discovery and exploration of non-invertible categorical symmetries~\cite{Brunner:2013xna,Bhardwaj:2017xup,Tachikawa:2017gyf,Thorngren:2021yso,Heidenreich:2021xpr,Choi:2021kmx,Kaidi:2021xfk,Roumpedakis:2022aik,Bhardwaj:2022yxj,Arias-Tamargo:2022nlf,Choi:2022zal,Antinucci:2022eat,Damia:2022rxw,Bhardwaj:2022lsg,Bartsch:2022mpm,Lin:2022xod,Kaidi:2022cpf,Decoppet:2022dnz,Bhardwaj:2022kot,Bartsch:2022ytj,Bhardwaj:2022maz,Delcamp:2023kew,Kaidi:2023maf,Radhakrishnan:2023zcq,Bhardwaj:2023wzd,Bartsch:2023pzl}. It is therefore natural to ask whether this method captures the entire landscape of categorical symmetries. In 1+1 and 3+1 dimensions, many non-invertible symmetries arise that cannot be constructed by finite gauging alone. However, there is strong reason to believe that 2+1 dimensions occupy a special position, where all categorical symmetries might be realized through generalized finite gauging.

This expectation is partially motivated by the sandwich construction of categorical symmetries~\cite{Freed:2012bs,Gaiotto:2020iye,Ji:2019jhk,Kong:2020cie,Freed:2022qnc,Freed:2022iao,Apruzzi:2021nmk}. On the one hand, 3+1 dimensional topological orders are tame and mostly correspond to Dijkgraaf-Witten gauge theories~\cite{Lan:2018vjb,Lan:2018bui,Johnson-Freyd:2020usu}. On the other hand, the gapped boundary conditions of these topological orders are far richer due to the diversity of 2+1 dimensional topological quantum field theories (TQFTs)~\cite{Wang2018GappedBT,Bullivant:2020xhy,PhysRevB.107.125425,Zhao:2022yaw,PhysRevB.94.045113,KONG201762}.

Further intuition arises from recent advances in the structure and classification of fusion 2-categories~\cite{2018arXiv181211933D,Johnson-Freyd:2020ivj,Decoppet2022TheMT,Decoppet2023Fiber2A,decoppet2024local}, which capture categorical symmetries in 2+1 dimensions. These developments, culminating in an upcoming homotopy classification~\cite{class-2fus}, suggest that all symmetries in 2+1 dimensions may be realized through suitably generalized gauging operations.

In this paper, we aim to systematically enumerate and classify gauging operations in 2+1 dimensional oriented quantum field theories with anomalous finite group symmetries. This classification is equivalent to classifying 2+1 dimensional oriented TQFTs or gapped systems\footnote{We use the terminology \emph{gapped systems} as opposed to \emph{gapped phases} to refer to isomorphism classes of topological quantum field theories as opposed to deformation classes. For example, our classification does not directly incorporate the 2+1 dimensional $E_8$-phase.} with anomalous finite group symmetries, or equivalently, the gapped boundary conditions for 3+1 dimensional Dijkgraaf-Witten theory.

We expect that this framework will encompass all finite categorical symmetries in 2+1 dimensional oriented quantum field theories.

\subsection{Summary of Results}

We examine finite symmetries in 2+1 dimensional oriented quantum field theories (QFTs), corresponding to spacetime symmetry type \( SO \) as defined in~\cite{Freed:2016rqq}. Our starting point is an oriented 2+1 dimensional QFT \( \mathcal{T} \) with a finite symmetry group \( G \) and a 't Hooft anomaly represented by
\be
\alpha \in Z^4(G,\mathbb{C}^\times) \, .
\ee
Even in the absence of symmetry background fields, \( \mathcal{T} \) may be viewed as the boundary of a 3+1 dimensional invertible theory whose partition function on a closed four-manifold \( M_4 \) is
\be
\zeta^{\sigma(M_4)} \, ,
\ee
where $\sigma(M_4)$ is the signature.
Although this is deformation-trivial and hence not a gravitational 't Hooft anomaly, tracking the parameter \( \zeta \) is important, as it may shift under the gauging operations we consider.\footnote{The 2+1 dimensional $E_8$-phase can be interpreted as a monodromy domain wall for $\zeta$.}

In general, the group \( G \) is anomalous and cannot be gauged directly. However, we can first stack \( \mathcal{T} \) with a 2+1 dimensional oriented topological quantum field theory (TQFT) equipped with a \( G \) symmetry and the opposite 't Hooft anomaly \( \alpha^{-1} \). Gauging the diagonal combination then yields a new theory, which we represent as:
\be
\cT \longrightarrow \cT \otimes \lambda \longrightarrow \cT /\!_\lambda\, G \, .
\ee
Here, \( \lambda \) may again bound a 3+1 dimensional invertible theory determined by \( \xi_\lambda \), and this gauging operation induces a shift \( \zeta \to \zeta \cdot \xi_\lambda \), where \( \xi_\lambda \) is the multiplicative central charge of \( \lambda \).

\begin{figure}[h]
\centering
\includegraphics[height=7cm]{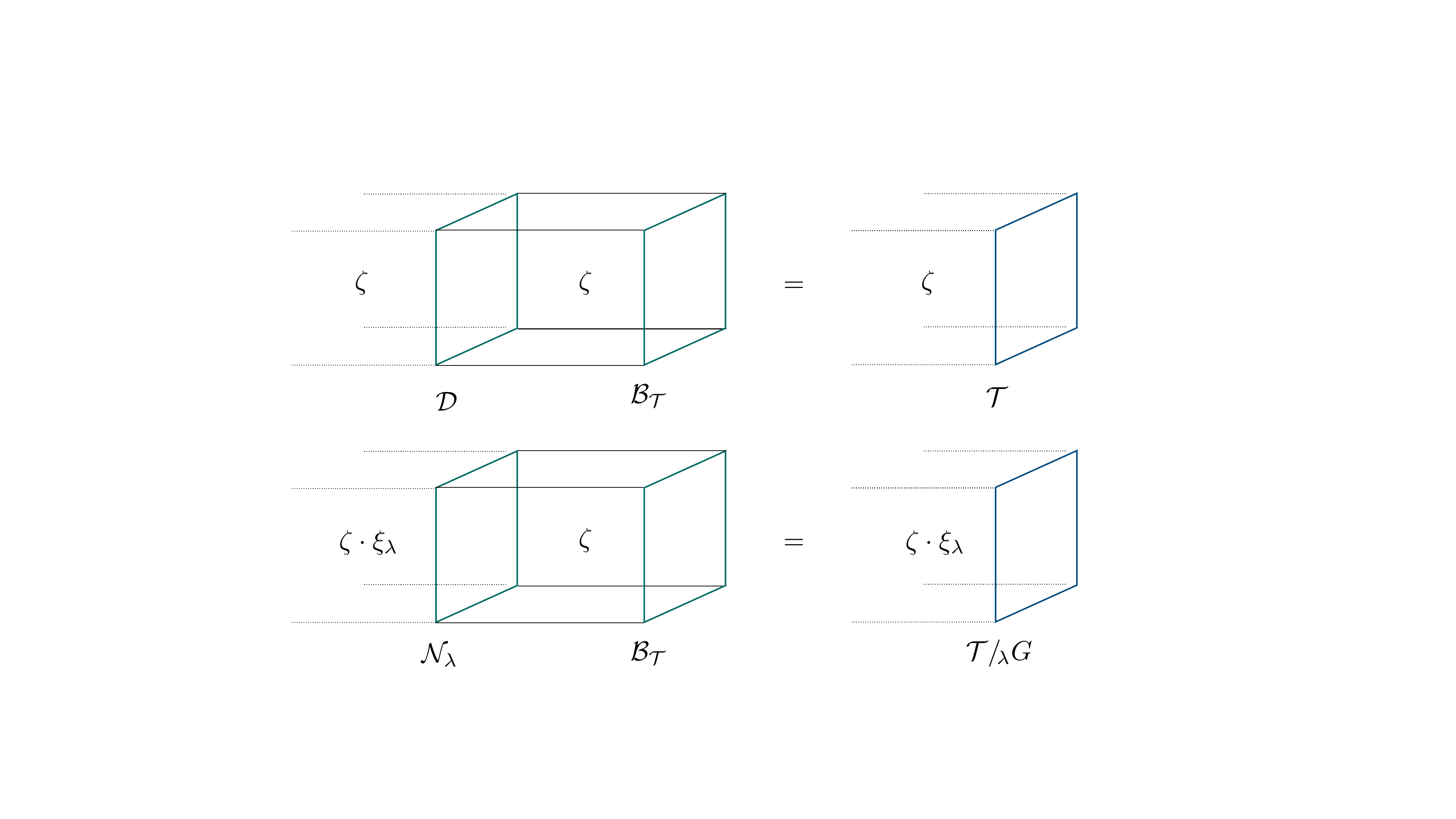}
\caption{Sandwich construction with Dirichlet and Neumann boundary conditions.}
\label{fig:intro-sandwich}
\end{figure}

This construction is visualized using the sandwich construction, as illustrated in figure~\ref{fig:intro-sandwich}. In this framework, the bulk is a 3+1 dimensional Dijkgraaf-Witten theory with symmetry \( G \) and action \( \alpha \), supplemented by the invertible theory \( \zeta \). Interval compactification with Dirichlet and Neumann boundary conditions results in:
\begin{itemize}
    \item A Dirichlet boundary condition \( \mathcal{D} \) yields the original theory \( \mathcal{T} \).
    \item A Neumann boundary condition \( \mathcal{N}_\lambda \) yields the gauged theory \( \mathcal{T} /_\lambda G \).
\end{itemize}
The Neumann boundary condition is constructed as above,
\be
 \mathcal{D} \longrightarrow \mathcal{D} \otimes \lambda \longrightarrow \mathcal{N}_\lambda := \mathcal{D} /_\lambda G  \, .
 \ee
This is really a gapped interface with an invertible theory, corresponding to a shift \( \zeta \to \zeta \cdot \xi_\lambda \) across the Neumann boundary condition.

Building on the result that all 3+1 dimensional oriented topological orders are Dijkgraaf-Witten theories~\cite{Lan:2018vjb,Lan:2018bui,Johnson-Freyd:2020usu}, we conclude that these operations generate all possible categorical symmetries in oriented 2+1 dimensional theories.

Our goal is to classify the data \( \lambda \). We start by considering a direct sum of 2+1 dimensional oriented TQFTs determined by a collection of modular fusion categories \( \{ \mathsf{b}_1, \ldots, \mathsf{b}_n \} \), and enumerate the data of a \( G \)-symmetry with 't Hooft anomaly \( \alpha^{-1} \). The symmetry \( G \) must act irreducibly to avoid decomposition, implying that \( \mathsf{b}_j \cong \mathsf{b} \) and \( G \) permutes the summands transitively.

This classification is summarized by:
\begin{enumerate}
    \item A subgroup \( H \subseteq G \).
    \item A modular fusion category \( \mathsf{b} \).
    \item An \( (\alpha|_H)^{-1} \)-twisted \( H \)-crossed extension of \( \mathsf{b} \).
\end{enumerate}
We introduce a generalisation of the classification of~\cite{Barkeshli_2019} from $H$-crossed extensions to $(\alpha|_H)^{-1}$-twisted $H$-crossed extensions. Notably, in this classification the subgroup \( H \subseteq G \) is not necessarily anomaly-free.

A more detailed classification is:
\begin{enumerate}
\item A subgroup $H \subseteq G$.
\item A modular fusion category $\mathsf{b}$.
\item A homomorphism $\rho : H \to \text{Aut}(\mathsf{b})$.
\item A 2-cochain $\psi \in C^2(H,\mathsf{b}^\times)$ satisfying $\delta\psi = o_3$.
\item A 3-cochain $\phi \in C^3(H,\bC^\times)$ satisfying $ \delta \phi = (\alpha|_H)^{-1} o_4 $.
\end{enumerate}
Here, \( \text{Aut}(\mathsf{b}) \) is the group of braided automorphisms of \( \mathsf{b} \), and \( o_3 \), \( o_4 \) are obstruction terms discussed in the main text.

Special cases include the trivial scenario \( \mathsf{b} = \mathsf{Vect} \), which leads to examples classified by subgroups \( H \subseteq G \) and trivializations \( \delta \phi = (\alpha|_H)^{-1} \), revisiting examples from our previous work~\cite{Bartsch:2022mpm,Bartsch:2022ytj}. This paper continues that exploration.

From a mathematical perspective, we introduce \( \alpha^{-1} \)-twisted \( G \)-crossed extensions of direct sums of modular fusion categories, which may be formulated as \( \alpha^{-1} \)-projective 3-representations of \( G \), and study the induction of irreducible 3-representations from subgroups \( H \subseteq G \).

Finally, the finite symmetries of a 2+1 dimensional oriented theory are encoded by a spherical fusion 2-category~\cite{Douglas:2018qfz}. We use the following notation for symmetry categories:
\begin{itemize}
    \item \( \mathsf{C} \) is the symmetry category of \( \mathcal{T} \) or the Dirichlet boundary condition \( \mathcal{D} \).
    \item \( \mathsf{C}_\lambda \) is the symmetry category of \( \mathcal{T} /_\lambda G \) or the Neumann boundary condition \( \mathcal{N}_\lambda \).
\end{itemize}
The former is given by:
\[
\mathsf{C} := \mathsf{2Vect}^\alpha[G] \, .
\]
The latter requires recognising that an $\alpha^{-1}$-twisted $G$-crossed extension determines a Lagrangian algebra, $L_\lambda \in \mathsf{Z}(\C)$, which is another description of gapped boundary conditions in 2+1 dimensional Dijkgraaf-Witten theory. The symmetry on the Neumann boundary condition is then given by

\[
\mathsf{C}_\lambda := \mathsf{Mod}_{\mathsf{Z}(\mathsf{C})}(L_\lambda) \, .
\]
This extends the group-theoretical fusion 2-categories studied in~\cite{Bartsch:2022mpm,Bhardwaj:2022lsg,Bartsch:2022ytj,Bhardwaj:2022maz,Decoppet2023Fiber2A}. While a full analysis is beyond the scope of this paper, we offer key observations and examples.

We propose that \( \mathsf{C}_\lambda \), with \( \lambda \) running over all classified data, generates all equivalence classes of spherical fusion 2-categories in oriented 2+1 dimensional quantum field theories, potentially recovering the announced mathematical classification of bosonic fusion 2-categories~\cite{class-2fus}, with \( \mathsf{b} \) interpreted as a non-degenerate braided fusion category.

\subsection{Future Directions}

An interesting avenue for future exploration is to extend the classification to different symmetry types, such as \( Spin \), \( O \), and \( Pin^\pm \). Additionally, the incorporation of unitarity is important for refining the classification. In the case of oriented theories, we expect the only modification to be the requirement that \( \mathsf{b} \) is a unitary modular fusion category.

Furthermore, applying the results of this work to gapped systems with categorical symmetry offers significant potential. Notable recent work in this area includes~\cite{Chang:2018iay, Komargodski:2020mxz, Huang:2021zvu, Inamura:2021szw, Inamura:2022lun, Bhardwaj:2023idu, Bhardwaj:2024qiv}. The sandwich construction suggests that the classification of gapped systems with \( \mathsf{C} \)-symmetry and those with \( \mathsf{C}_\lambda \)-symmetry in 2+1 dimensions should coincide, as both are governed by the same 3+1 dimensional bulk theory. However, this correspondence is challenging to interpret physically, due to the intricate Morita equivalence between \( \mathsf{C} \) and \( \mathsf{C}_\lambda \). 

Thus, a more direct and physically transparent classification scheme is desirable. We anticipate that the methods developed in this paper will be helpful in tackling this problem and providing new insights into the classification of gapped systems with categorical symmetry in 2+1 dimensions.

\vspace{10pt}

\emph{Note: during the preparation of this paper, we became aware of~\cite{Bhardwaj:2024qiv}, which adopts a similar approach to investigate 2+1 dimensional gapped systems with non-invertible symmetries. However, their analysis is restricted to anomaly-free subgroups \( H \subseteq G \) and (untwisted) \( H \)-crossed extensions of modular fusion categories.}



\section{Warm-up: 1+1 dimensions}
\label{sec:2d}

While our primary focus is 2+1 dimensions, we first review analogous constructions in 1+1 dimensions to provide some motivation for the strategy used in section~\ref{sec:3d}.

\subsection{Symmetries and Anomalies}

The finite symmetries of a 2-dimensional oriented theory are captured by a spherical fusion category. The spherical fusion structure ensures the associated 3-dimensional TQFT in the sandwich construction is fully-extended and oriented.\footnote{A precise statement is a spherical fusion category is an $SO_3$ homotopy fixed point~\cite{Douglas:2013aea}. It possible to entertain a weaker requirement of pivotal fusion category or $SO_2$ homotopy fixed point or requirement that the symmetry TFT is combed. We will not do so here.}

Let us now suppose $\cT$ has a finite symmetry group $G$. 
The `t Hooft anomalies are classified by
\be
[\alpha] \in H^3(G,\bC^\times) 
\ee
and by choosing an appropriate local counter-term for background fields the `t Hooft anomaly can be represented by a normalised 3-cocycle $\alpha \in Z^3(G,\mathbb{C}^\times)$. We denote the symmetry structure by the combination $(G,\alpha)$.

\begin{figure}[h]
\centering
\includegraphics[height=3cm]{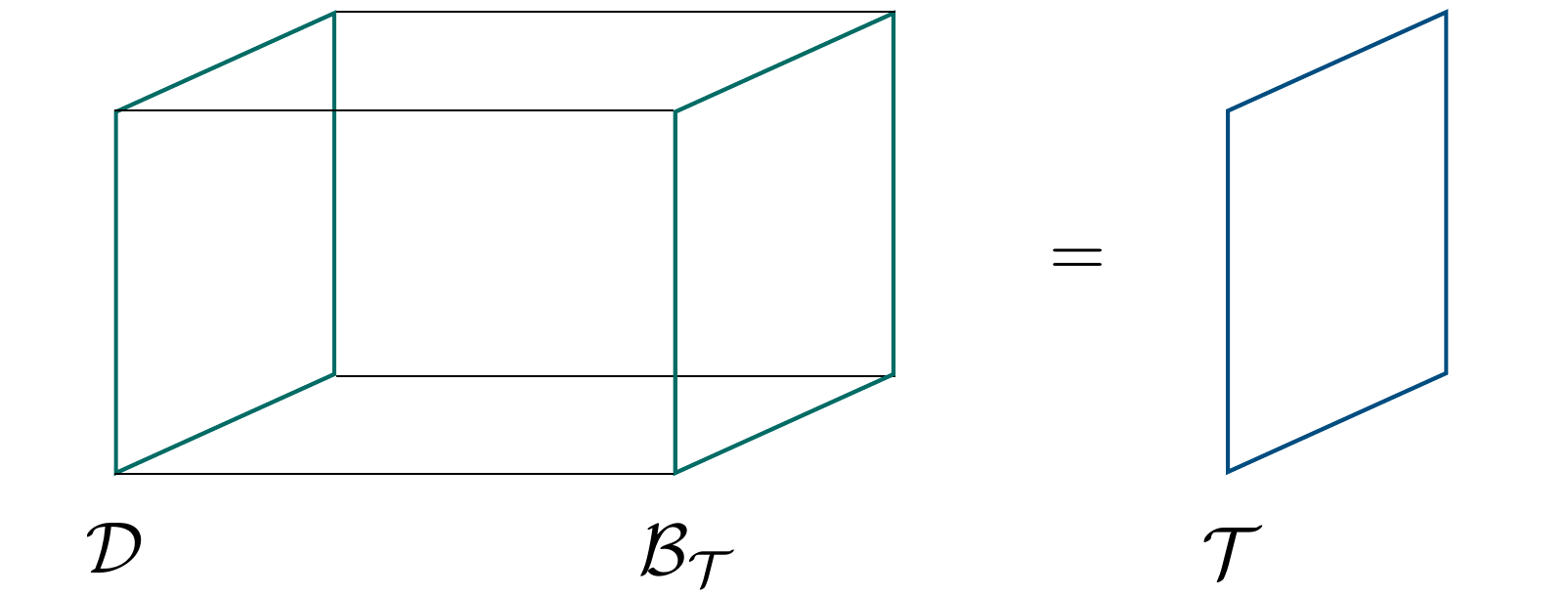}
\caption{Sandwich construction with Dirichlet boundary condition.}
\label{fig:2d-sandwich-1}
\end{figure}

The symmetry category is 
\be
\mathsf{C} = \mathsf{Vect}^\alpha[G] 
\ee
and the symmetry TFT is the corresponding 3-dimensional oriented Dijkgraaf-Witten theory. A 2-dimensional theory $\cT$ with this symmetry is obtained by interval compactification with a Dirichlet boundary condition $\cD$, as shown in figure~\ref{fig:2d-sandwich-1}.


\subsection{Gauging and Gapped Boundaries}

In general $G$, is anomalous and cannot be gauged without additional ingredients. However, we can first stack with a 2-dimensional oriented TQFT $\lambda$ with symmetry $(G,\alpha^{-1})$ and then gauge the anomaly free diagonal combination. 

\begin{figure}[h]
\centering
\includegraphics[height=3cm]{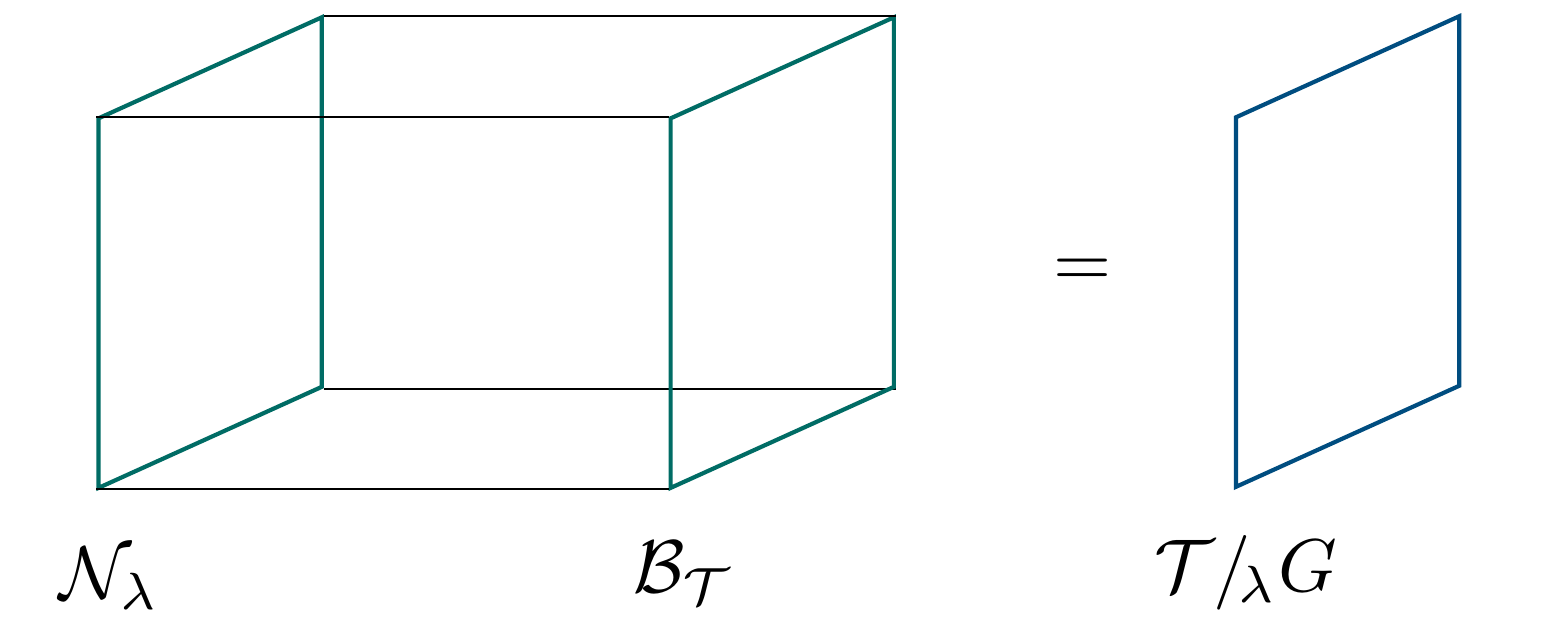}
\caption{Sandwich construction with Neumann boundary condition.}
\label{fig:2d-sandwich-2}
\end{figure}

We summarise these steps by
\be
\cT \longrightarrow \cT \otimes \lambda \longrightarrow \cT /\!_\lambda\, G \, .
\ee
In the sandwich construction, $\cT /\!_\lambda\, G$ is recovered by interval compactification with a Neumann boundary condition $\cN_\lambda$ defined by an analogous operation
\be
\cD \longrightarrow \cD \otimes \lambda \longrightarrow \cN_\lambda := \cD /\!_\lambda\, G  \, .
\ee
This is shown in figure~\ref{fig:2d-sandwich-2}.

We must therefore classify 2-dimensional oriented TQFTs $\lambda$ with symmetry $(G,\alpha^{-1})$. From a mathematical perspective, this can be formulated in terms of $G$-crossed extensions or projective 2-representations of $G$. 


\subsection{Formulation as Crossed Extensions}

First, a 2-dimensional oriented TQFT without a global symmetry is determined by a finite-dimensional commutative Frobenius algebra $C_e$ of topological local operators. This must take the form\footnote{The reason for the subscript $e$ will become clear momentarily.}
\be
C_e  = \bigoplus_{j=1}^n \bC \, ,
\label{eq:2d-comm-C*}
\ee
with Frobenius structure
\bea
\text{Tr}
& : C_e \to \bC \\
& : \{c_1,\ldots,c_n\} \mapsto \sum_{j=1}^n t_jc_j
\eea
determined by non-zero complex numbers $t_1,\ldots,t_n \in \mathbb{C}^\times$. The numbers are Euler counter-term contributions.\footnote{A fully-extended 2-dimensional oriented theory is specified a finite-dimensional Frobenius algebra $A$ of topological local operators on a regular boundary condition. This has $Z(A) = C$ and Frobenius structure determined by parameters $\lambda_j \in \mathbb{C}^\times$ with $\lambda_j^2 = t_j$. For our purposes, it is enough to work with the bulk data $C$, $t_j$.}

Compatibility with the structure of a spherical fusion category will require 
\be
t_i = t_j 
\ee
for all $i \neq j$. If not, gauging will result in a multi-fusion structure with relative Euler terms that shift the pivotal structure away from spherical. From a symmetry TFT perspective, this condition should ensure it is consistent as a gapped boundary condition for a trivial 3-dimensional oriented TQFT.\footnote{A commutative Frobenius algebra can be understood as a commutative algebra in $\mathsf{Vect}$, viewed as a spherical braided fusion category ($SO_2$ homotopy fixed point), but this is only compatible with $\mathsf{Vect}$ as a modular fusion category ($SO_3$ homotopy fixed point) if it is special.} This leaves an overall Euler term $t \in \bC^\times$ that we set to unity for convenience.

The data needed to equip this theory with a $(G,\alpha^{-1})$ symmetry was determined in~\cite{moore2006d}.  The construction incorporates not just genuine local operators but twisted sector operators attached to topological lines labelled by group elements $g \in $G. We will formulate it here as a $\alpha^{-1}$-twisted $G$-crossed extension of $C_e$.

\begin{figure}[h]
\centering
\includegraphics[height=2.75cm]{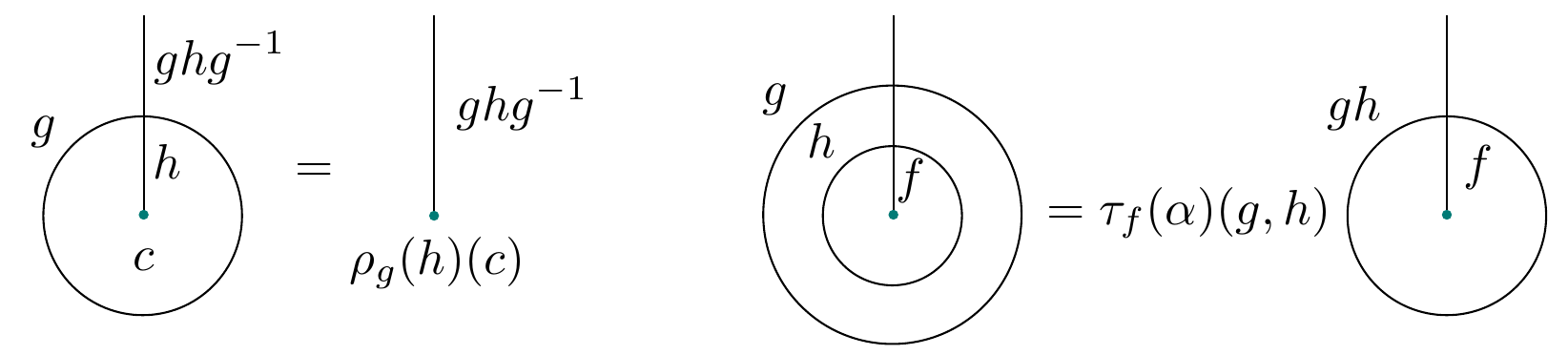}
\caption{Illustration of linear maps $\rho_{g}(h) : C_h \to C_{ghg^{-1}}$ and twisted composition.}
\label{fig:2d-tube}
\end{figure}

The first step consists of:
\begin{enumerate}
\item A finite-dimensional $G$-graded vector space $C = \bigoplus_{g \in G} C_g$.
\item A $G$-action by invertible linear maps $\rho_{g}(h) : C_h \to C_{ghg^{-1}}$ satisfying twisted composition law
	\be
	\rho_{g}({}^hf) \circ \rho_{h}(f) =\tau_f(\alpha)(g,h) \rho_{gh}(f) \, 
	\label{eq:2d-twisted-composition}
	\ee
	for all $g,h,f \in G$.
\end{enumerate}
The collections of phases
\be
\tau_f(\alpha)(g,h) := \frac{\alpha(g,{}^hf,h)}{\alpha({}^{gh}f,g,h)\alpha(g,h,f)}
\label{eq:2d-transgression}
\ee
defines a groupoid 2-cocycle
\be
\tau(\alpha) \in Z^2(G /\!/G,U(1)) 
\ee
or equivalently a collection of group 2-cocycles $\tau_f(\alpha) \in Z^2(C_f(G),\mathbb{C}^\times)$ by restricting to arguments in the centralizer $g,h \in C_f(G)$. This is the transgression of $\alpha$. 
The interpretation of this data is illustrated in figure~\ref{fig:2d-tube}. 

\begin{figure}[h]
\centering
\includegraphics[height=5cm]{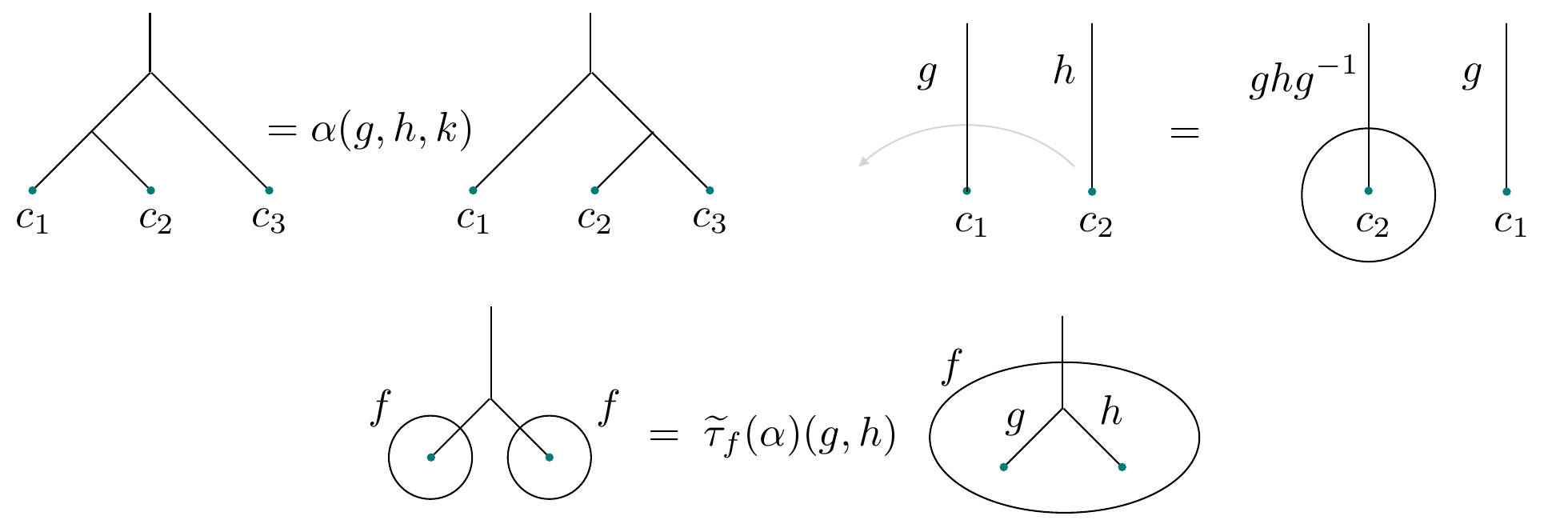}
\caption{Twisted associativity, crossed commutativity and twisted distributivity.}
\label{fig:2d-tube-2}
\end{figure}

The second step incorporates a graded multiplication $C_g C_h \subseteq C_{gh}$ that satisfies:
\begin{enumerate}
	\item Twisted associativity: 
	\be
	(c_1 c_2) c_3 = \alpha(g,h,k) c_1 (c_2c_3)
	\ee
	for all $c_1 \in C_g$, $c_2 \in C_h$, $c_3 \in C_k$.
	\item Crossed commutativity: 
	\be
	c_1 c_2 = \rho_{g}(h)(c_2) c_1
	\ee
	for all $c_1 \in C_g$, $c_2 \in C_h$.
	\item Twisted distributivity:
	\be
	\rho_{f}(g)\rho_{f}(h) = \widetilde{\tau}_f(\alpha)(g,h)\rho_f(gh)
	\label{eq:2d-twisted-distributivity}
	\ee
	for all $f,g,h \in G$.
\end{enumerate}
The latter exhibits another collection of phases
\be
\widetilde{\tau}_f(\alpha)(g,h) := \frac{\alpha({}^fg, f, h)}{\alpha(f,g,h)\alpha({}^fg,{}^fh,f)}
\ee
satisfying the same properties as the transgression. The Frobenius structure on $C_e$ extends to $\text{Tr} : C \to \bC
$ by zero on non-identity components. The interpretation of the multiplicative structure is illustrated in figure~\ref{fig:2d-tube-2}. 

We call $C$ an $\alpha^{-1}$-twisted $G$-crossed extension of $C_e$. The maps $\rho_g(e) : C_e \to C_e$ permute the summands in $C_e$ and determine a permutation representation 
\be
\sigma : G \to  S_n \, .
\ee
We say $C$ is irreducible if this action permutes the summands transitively, or equivalently, if the permutation representation is irreducible. 

A convenient way to summarise the above data is that an irreducible $\alpha^{-1}$-twisted $G$-crossed extension defines a Lagrangian algebra $L_\lambda \in \mathsf{Z}(\C)$ in the Drinfeld center of the symmetry category of the Dirichlet boundary condition $\cD$. This is precisely the Lagrangian algebra corresponding to the Neumann boundary condition $\cN_\lambda$.


\subsection{Formulation as 2-representations}

We now explain how such extensions $C$ are in 1-1 correspondence with $\alpha^{-1}$-projective 2-representations of $G$ on $C_e$. For mathematical background on 2-representation theory and derivations of the results presented below we refer to~\cite{elgueta2005representation,ganter2007representation,bartlett2009unitary,OSORNO2010369,GANTER2015301,ganter2017representation}. 

With a eye towards even higher representations in the next section, we will view a $\alpha^{-1}$-projective 2-representation is a 2-group homomorphism
\be
\lambda : \mathsf{G}_\alpha \longrightarrow \mathsf{Aut}(C_e) \, ,
\ee
where:
\begin{itemize}
\item The source is the 2-group extension with homotopy groups
\bea
\pi_0(\mathsf{G}_\alpha) & = G \\
\pi_1(\mathsf{G}_\alpha) & = \bC^\times
\eea
and Postnikov invariant $\alpha^{-1} \in H^3(G, \bC^\times)$. 
\item The target is the automorphism 2-group capturing invertible symmetries and 't Hooft anomalies of the associated 2-dimensional oriented TQFT. This has homotopy groups 
\bea
\pi_0(\mathsf{Aut}(C_e)) & =  S_n \\
\pi_1(\mathsf{Aut}(C_e)) & =  (\bC^\times)^n 
\eea
where the first permutes the summands of $C_e$ and acts on the second by permutations. The Postnikov invariant is trivial and the automorphism 2-group is split. 
\end{itemize}

It is now possible to enumerate the $\alpha^{-1}$-projective 2-representations explicitly. They are determined by:
\begin{enumerate}
	\item A permutation representation $\sigma : G \to S_n$.
	\item A 2-cochain $c \in C^2(G,(\mathbb{C}^\times)^n)$ satisfying $\delta c = \alpha^{-1}$.
\end{enumerate}
The latter is a collection of maps 
\be
c_j : G \times G \to \bC^\times
\ee
satisfying the conditions
\be\label{eq:c-triv-alpha}
\frac{c_{\sigma_g^{-1}(j)}(h,k)c_j(g,hk)}{c_j(gh,k)c_j(g,h)} =  \alpha(g,h,k)^{-1}
\ee
for all $g,h,k \in G$ and  $j = 1,\ldots,n$. The projective 2-representation is irreducible if the permutation representation $\sigma : G \to S_n$ is irreducible.

The associated $G$-crossed extension of $C_e$ is given by
\be
C_{g} = \bigoplus_{j \, | \, \sigma_g(j) = j } \mathbb{C} \cdot e_g^j
\ee
with generators $e_g^j$ with $\sigma_g(j) = j$ satisfying
\be
\rho_{g}(h)  e^j_h  = \frac{c_{\sigma_g(j)}(g, h)}{c_{\sigma_g(j)}({}^gh, g)} e^{\sigma_g(j)}_{ghg^{-1}} \, , \qquad e^j_g \cdot e^j_h = c_j(g,h) e^j_{gh} \, .
\ee
This provides an explicit parametrisation of $\alpha^{-1}$-twisted $G$-crossed extensions of $C_e$.


\subsection{Subgroups and Induction}

An irreducible projective 2-representation $\lambda$ has an equivalent description as
\begin{enumerate}
	\item A subgroup $H \subseteq G$.
	\item A 2-cochain $\psi \in C^2(H,\bC^\times)$ satisfying $ \delta \psi = (\alpha|_H)^{-1}$.
\end{enumerate}
Concretely, the 2-cochain is a map $\psi : H \times H \to U(1)$ satisfying
\be
\frac{\psi(h_2,h_3)\psi(h_1,h_2h_3)}{\psi(h_1h_2,h_3)\psi(h_1,h_2)} =  \alpha(h_1,h_2,h_3)^{-1}
\label{eq:2d-triv-def}
\ee
for all $h_1,h_2,h_3 \in H$.

This data determines an irreducible projective 2-representation by induction. First, choosing coset representatives $a_j H$ determines a permutation representation $\sigma$ on $G/H \cong \{a_1,\ldots,a_n\}$ by 
\be
g \cdot a_j H = a_{\sigma_g(j)} H
\ee
with compensating transformations
\be
\ell_{g,j} = a_{\sigma_g(j)}^{-1} \cdot g \cdot a_j \, .
\ee
The 2-cochain $\psi$ induces a 2-cochain $\{c_1,...,c_n\}\in C^2(G, (\bC^\times)^{G/H})$ that trivialises $\alpha^{-1}$ as in equation~\eqref{eq:c-triv-alpha} and takes the form
\be
c_j(g_1,g_2) \propto \psi(\ell_{g_1,\sigma^{-1}_{g_1}(j)},\ell_{g_2,\sigma^{-1}_{g_1g_2}(j)}) \, 
\ee
where we have omitted a factor that is independent of $\psi$.
It is known that all irreducible projective 2-representations arise in this manner. Further background on induction of 2-representations can be found in~\cite{ganter2007representation,bartlett2009unitary,OSORNO2010369}.

The associated $\alpha^{-1}$-twisted $G$-crossed extension is similarly determined by induction. It has graded components
\be
C_{g} = \bigoplus_{i \, | \, g \in {}^{a_i}H} \mathbb{C} \cdot e^{i}_g \, ,
\ee
where the generators $e^j_g$ with $g \in {}^{a_j}H$ satisfy 
\be\label{eq:2d-induction}
\rho_{g}(h)  e^j_h = \frac{c_j(g, h)}{c_j({}^gh, g)}    e^{\sigma_g(j)}_{ghg^{-1}} \qquad \qquad e^j_{h_1} \cdot e^j_{h_2} = \psi^{a_j}(h_1,h_2) e^j_{h_1h_2}
\ee
and $\psi^{a_j}\in Z^2({}^{a_j}H,\bC^\times)$ is defined by
\be
\psi^{a_j}(h_1,h_2) = \psi(h_1^{a_j},h_2^{a_j}) 
\ee
for all $h_1,h_2 \in {}^{a_j}H$, where it overlaps with $c_j(h_1, h_2)$. The provides another parametrisation of irreducible $G$-crossed extensions of $C_e$.

The resulting gauging operation is equivalent to gauging the subgroup $H \subseteq G$ with a local counter-term $\psi$ trivialising the anomaly $(\alpha|_H)^{-1}$. The gauging operation $\cT  \longrightarrow \cT /\!_\lambda\, G$
can thus equivalently be represented by
\be
\cT  \longrightarrow \cT /\!_\psi\, H  \, .
\ee
There is no physical distinction between gauging $H \subset G$ and ${}^a H \subset G$, which generate equivalent projective 2-representations and twisted $G$-crossed extensions.

\subsection{Symmetry Categories}

Let us now consider the symmetry category of the gauged theory $\cT /\!_\psi\, H$. The symmetry category supported on the  Neumann boundary condition $\cN_\lambda$ is determined by computing modules over the associated Lagrangian algebra,
\be
\mathsf{C}_{\lambda} := \mathsf{Mod}_{\mathsf{Z}(\C)}(L_\lambda) \, .
\ee
The outcome,
\be
\mathsf{C}_{\lambda} \cong \mathsf{C}(G,\alpha | H ,\psi) \, ,
\ee
is a group-theoretical fusion category. The structure of group-theoretical fusion categories has been studied extensively in the mathematical literature~\cite{ostrik2006module,etingof2005fusion,GELAKI20092631,10.1063/1.4774293} and explored in the context of categorical symmetries in 1+1 dimensions in~\cite{Bartsch:2022mpm,Bartsch:2022ytj}. 

Let summarise briefly summarise some features that are duplicated in 2+1 dimensions. First, there is an inclusion
\be
\mathsf{Rep}(H) \hookrightarrow \mathsf{C}_{\lambda} 
\ee 
generated by topological Wilson lines for $H$. Second, it admits a direct sum decomposition as a finite semi-simple category,
\be
\mathsf{C}_{\lambda} \cong \bigoplus_{[g] \in H\backslash G /H} \mathsf{Rep}^{c_g}(H \cap {}^g H) \, ,
\ee
where the summation is over double cosets with representatives $g \in G$ with summands given by projective representations of $H \cap {}^gH$ with 2-cocycle $c_g \in Z^2(H \cap {}^g H,\mathbb{C}^\times)$ built from $\alpha$, $\psi$. The fusion rules build upon the foundation of the double coset ring supplemented with instructions for decomposing and combining projective representations.


\section{Main event: 2+1 dimensions}
\label{sec:3d}

\subsection{Symmetries and Anomalies}

The finite symmetries of a three-dimensional oriented theory $\cT$ are captured by a spherical fusion 2-category $\C_\cT$. The spherical structure ensures, amoung other things, a well-defined notion of dimension from topological surfaces wrapped on spheres and that the associated 4-dimensional symmetry TFT is fully-extended and oriented.\footnote{A spherical fusion 2-category is expected to correspond to an $SO_4$ homotopy fixed point~\cite{Douglas:2018qfz}.} For mathematical background and definitions see~\cite{Douglas:2018qfz}.

\begin{figure}[h]
\centering
\includegraphics[height=3cm]{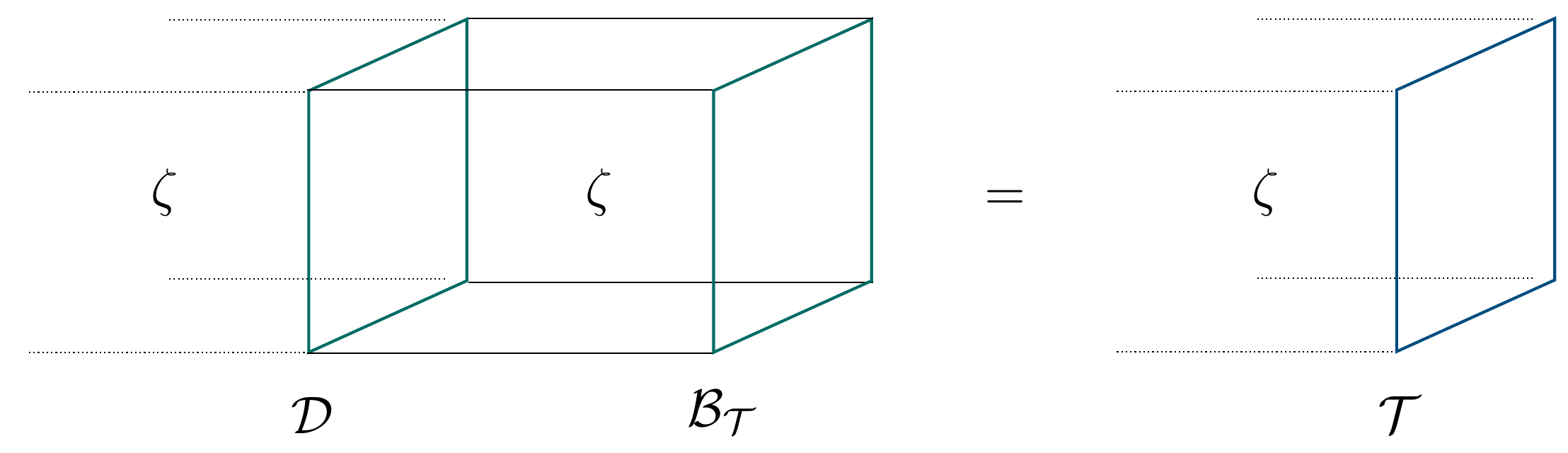}
\caption{Sandwich construction with Neumann boundary condition.}
\label{fig:3d-sandwich-1}
\end{figure}

Let us now suppose $\cT$ has a finite symmetry group $G$. 
The `t Hooft anomalies are classified by group cohomology
and by choosing an appropriate local counter-term in background fields it can be represented by a normalised 4-cocycle 
\be
\alpha \in Z^4(G,\mathbb{C}^\times)\, .
\ee
The associated spherical fusion 2-category is
\be
\mathsf{C} = \mathsf{2Vect}^\alpha[G] \, .
\ee
The same data determines a 4-dimensional oriented Dijkgraaf-Witten theory used in the sandwich construction, in which the symmetry $(G,\alpha)$ is realised on the regular Dirichlet boundary condition. A 3-dimensional theory $\cT$ with symmetry $(G,\alpha)$ is obtained from an interval compactification, as shown in figure~\ref{fig:3d-sandwich-1}.

Even without background fields for the global symmetry, $\cT$ may sit at the boundary of a 4-dimensional invertible theory whose partition function on a closed oriented four-manifold $M_4$ is
\be
\zeta^{\sigma(M_4)} \, ,
\ee
where $\sigma(M_4)$ is the signature and $\zeta \in \bC^\times$. This is deformation trivial so we refrain from calling it a gravitational anomaly. Nevertheless, $\zeta$ may shift under the gauging operations we consider and therefore it is useful to keep track of it.

The parameter $\zeta$ is incorporated in the sandwich construction by stacking 4-dimensional Dijkgraaf-Witten theory with invertible theories and relaxing gapped boundary conditions to gapped interfaces with invertible theories. This is illustrated in figure.

\subsection{Gauging and Gapped Boundaries}

In general, $G$ is anomalous and cannot be gauged without additional ingredients. However, we can always stack with a 3-dimensional oriented TQFT $\lambda$ with symmetry $(G,\alpha^{-1})$ and then gauging the anomaly free diagonal combination. 

We summarise this by
\be
\cT \longrightarrow \cT \otimes \lambda \longrightarrow \cT /\!_\lambda\, G \, .
\label{eq:3d-steps}
\ee
Note $\lambda$ need not be a genuine 3-dimensional oriented theory but a boundary condition of a 4-dimensional oriented invertible theory with parameter $\zeta_\lambda$. The resulting gauging operation~\eqref{eq:3d-steps} will shift
\be
\zeta_{\cT /\!_\lambda\, G} =   \zeta_\cT \cdot \xi_\lambda
\ee
where $\xi_\lambda$ is the multiplicative central charge of $\lambda$. The multiplicative central charge is related to the chiral central charge by $\xi_\lambda = e^{2\pi c_\lambda/8}$.

\begin{figure}[h]
\centering
\includegraphics[height=3cm]{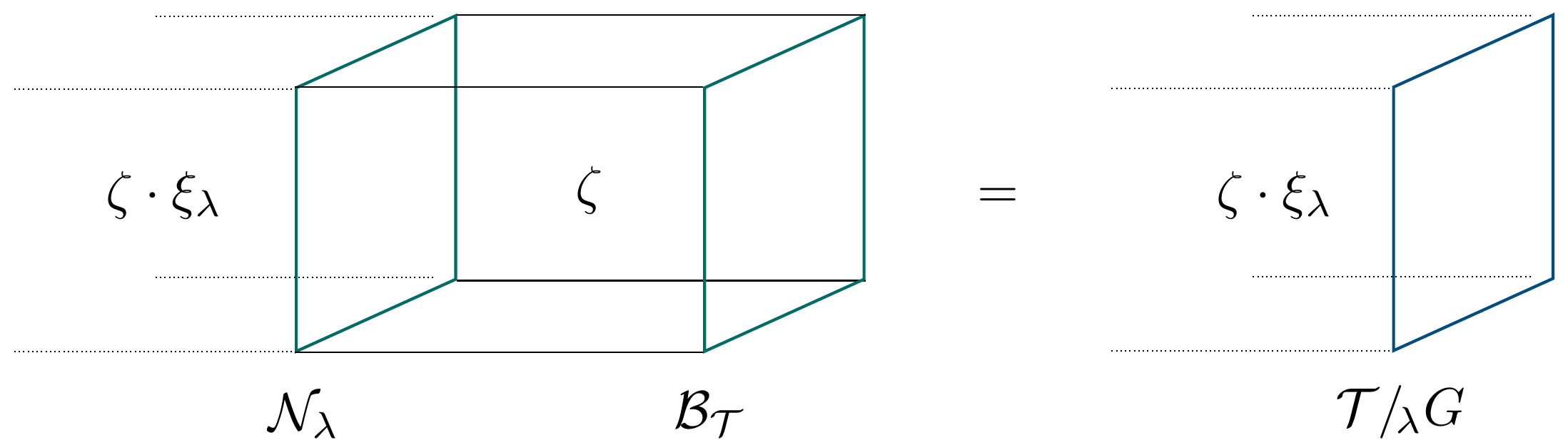}
\caption{Sandwich construction with Neumann boundary condition.}
\label{fig:3d-sandwich-2}
\end{figure}

In the sandwich construction, $\cT /\!_\lambda\, G$ is obtained by interval compactification with a Neumann boundary condition $\cN_\lambda$ formed by stacking the Dirichlet boundary condition with $\lambda$ and gauging $G$ on the boundary,
\be
\cD \longrightarrow \cD \otimes \lambda \longrightarrow \cN_\lambda:= \cD /\!_\lambda\, G \, .
\ee
Incorporating the parameter $\zeta$, the Neumann boundary condition $\cN_\lambda$ becomes a gapped interface across which 
\be
\zeta \to \zeta \cdot \xi_\lambda \, .
\ee
This is shown in figure~\ref{fig:3d-sandwich-2}.
It can be viewed as a gapped boundary condition of 4-dimensional Dijkgraaf-Witten theory as a deformation class, or up to invertible theories. We expect that this construction exhausts such gapped boundary conditions.

We must therefore classify 3-dimensional oriented TQFTs or gapped systems $\lambda$ with symmetry $(G,\alpha^{-1})$.\footnote{We use the terminology of gapped \emph{systems} as opposed to gapped \emph{phases} corresponding to isomorphism classes of 3-dimensional oriented TQFTs as opposed to deformation classes. The difference here is the $E_8$-phase. From the perspective of this paper, the $E_8$-phase is instead understood as a monodromy domain wall for the 4-dimensional invertible theory $\zeta^{\sigma(M_4)}$.} For those involving spontaneous symmetry breaking to an anomaly-free subgroup $H \subseteq G$ and supplemented with a $H$-symmetry gapped system, this reduces to $H$-crossed extensions of modular fusion categories~\cite{Barkeshli_2019}. 
We propose a generalisation to arbitrary subgroups $H \subseteq G$ and $(\alpha|_H)^{-1}$-twisted $H$-crossed extensions of modular fusion categories.

From a mathematical perspective, we introduce $\alpha^{-1}$-twisted $G$-crossed extensions of direct sums of modular fusion categories, which may be formulated in terms of $\alpha^{-1}$-projective 3-representations of $G$, and study induction from subgroups $H \subset G$.


\subsection{Formulation as Crossed Extensions}
\label{subsec:3d-crossed}

We begin by considering a direct sum of 3-dimensional oriented TQFTs before explaining how to equip this with symmetry structure $(G,\alpha^{-1})$.

A (not necessarily irreducible) 3-dimensional oriented TQFT is determined by
\be
\mathsf{B}_e = \bigoplus_{j=1}^n \mathsf{b}_j
\label{eq:braided-decomposition}
\ee
where each summand is spherical non-degenerate braided or equivalently modular fusion category. This should be compared with the decomposition~\eqref{eq:2d-comm-C*} in two dimensions. In contrast, here each summand has internal structure admitting braided automorphisms, which results in more interesting phenomena.

A more precise statement is that each summand $\mathsf{b}_j$ in equation~\eqref{eq:braided-decomposition} determines a 4-dimensional fully-extended invertible oriented TQFT with partition function
\be
\xi_j^{\sigma(M_4)}
\ee
where $\xi_j$ is the multiplicative central charge. The modular fusion category $\mathsf{b}_j$ then captures line defects on a gapped boundary condition and can be regarded as a 3-dimensional oriented TQFT up to a deformation-trivial or invertible 4-dimensional theory.

We must now equip this with symmetry structure $(G,\alpha^{-1})$. This should incorporate not only genuine line defects but also twisted sector line defects attached to surfaces indexed by $g \in G$.

The first step consists of introducing:
\begin{enumerate}
	\item A $G$-graded finite semi-simple category $\mathsf{B} = \bigoplus_{g \in G}\mathsf{B}_g$.
	\item A $G$-action by functors $\rho_g(h) : \mathsf{B}_h \to \mathsf{B}_{ghg^{-1}}$.
	\item Natural isomorphisms $\rho^{\circ}_{g,h}(k) : \rho_{g}({}^hk)\circ\rho_{h}(k) \rightarrow \rho_{gh}(k)$ satisfying the twisted tetrahedron relation
	\be
	\rho^\circ_{g, hk} (f)\rho^\circ_{h,k}(f) = \tau_f\alpha(g,h,k)\; \rho^\circ_{gh,k}(f) \rho^\circ_{g,h}({}^kf)
	\ee
	for all $g,h,k,f \in G$.
\end{enumerate}
The collection of phases
\be
\tau_f(\alpha)(g,h,k) = \frac{\alpha(g,h,{}^kf,k)\alpha({}^{ghk}f,g,h,k)}{\alpha(g,h,k,f)\alpha(g,{}^{hk}f,h,k)}
\label{eq:3d-transgression}
\ee
indexed by $f\in G$ define a groupoid 3-cocycle
\be
\tau(\alpha) \in Z^3(G/\!/G, \bC^\times)
\ee
or equivalently, a collection of 3-cocycles $\tau_f(\alpha)\in Z^3(C_f(G), \bC^\times)$ upon restriction to elements $g,h,k\in C_f(G)$. This is the transgression of $\alpha$.

The interpretation of this data is that $\B_g$ captures twisted sector topological lines attached to the symmetry defect $g \in G$, while the functors $\rho_g(h)$ and natural transformations $\rho^{\circ}_{g,h}(k)$ capture the consistent wrapping by cylindrical symmetry defects. We spare the reader our attempt at an illustration.\footnote{This data is determines a 2-representation of the tube 2-algebra of the symmetry category $\C = \mathsf{2Vect}^{\alpha}[G]$ introduced in~\cite{Bartsch:2023wvv} and relevant figures can be found therein.}

The second step incorporates a graded fusion 
\be
\mathsf{B}_g\otimes\mathsf{B}_h\subseteq\mathsf{B}_{gh}
\ee
equipped with the following structure:
\begin{enumerate}
	\item Twisted associativity: associator isomorphisms $c(b_1,b_2,b_3) : (b_1\otimes b_2)\otimes b_3\to b_1\otimes(b_2\otimes b_3)$ satisfying
	\be
	c(b_1b_2, b_3, b_4) c(b_1, b_2, b_3b_4) = \alpha(g,h,k,l)\; c(b_2,b_3,b_4) c(b_1, b_2b_3, b_4) c(b_1,b_2,b_3) 
	\ee
	for all $b_1\in\mathsf{B}_g$, $b_2\in\mathsf{B}_h$, $b_3\in\mathsf{B}_k$, $b_4\in\mathsf{B}_l$.
	\item Twisted distributivity: distributor isomorphisms $\rho^\otimes_f(g,h) : \rho_f(g)\otimes\rho_f(h)\to \rho_f(gh)$ satisfying
	\be
	\rho^\otimes_{f}(gh,k)\rho^\otimes_{f}(g,h)= \widetilde{\tau}_f\alpha(g,h,k)\; \rho^\otimes_{f}(g,hk)\rho^\otimes_{f}(h,k)
	\ee
	for all $g,h,k,f\in G$.
	\item Crossed braiding: half-braiding isomorphisms $\beta(b_1,b_2) : b_1\otimes b_2\to \rho_g(b_2)\otimes b_1$ satisfying
	\be
	\beta(b_1, b_2b_3) = \rho^\otimes_g(h,k)\frac{c(\rho_g(b_2), b_1, b_3)}{c(b_1,b_2, b_3)c(\rho_g(b_2), \rho_g(b_3), b_1)}\; \beta(b_1, b_2)\beta(b_1, b_3)
	\ee
	and
	\be
	\beta(b_1b_2,b_3) = \rho^\circ_{g,h}(k)\frac{c(b_1,b_2,b_3)c(\rho_{gh}(b_3), b_1, b_2)}{c(b_1, \rho_h(b_3), b_2)}\beta(b_2, b_3)\beta(b_1, \rho_h(b_3))
	\ee
	for all $b_1\in\mathsf{B}_g$, $b_2\in \mathsf{B}_h$, $b_3\in\mathsf{B}_k$.
\end{enumerate}
The twisted distributivity introduces a second set of phases
\be
\widetilde{\tau}_f(\alpha)(g,h,k) = \frac{\alpha({}^fg,f,h,k)\alpha({}^fg,{}^fh,{}^fk,f)}{\alpha(f,g,h,k)\alpha({}^fg,{}^fh,f,k)}
\ee
that satisfy the same cocycle condition as the transgression~\eqref{eq:3d-transgression}.

We call this data an $\alpha^{-1}$-twisted $G$-crossed extension of $\mathsf{B}_e$. Let us now discuss what it means for this extension to be irreducible. First, the functors $\rho_g(e) : \mathsf{B}_e \to \mathsf{B}_e$ are weakly invertible and determine a group homomorphism
\be
\Sigma : G \to \text{Aut}(\mathsf{B}_e)
\label{eq:3d-group-hom}
\ee
where $\text{Aut}(\mathsf{B}_e)$ is the group of braided automorphisms of $\mathsf{B}_e$. We say that an $\alpha^{-1}$-twisted $G$-crossed extension is irreducible if
\begin{enumerate}
\item the summands are equivalent $\mathsf{b}_j \cong \mathsf{b}$ for all $j = 1,\ldots,n$, and 
\item the group homomorphism $\sigma : G \to S_n$ obtained by restricting to the normal subgroup $S_n \subset \text{Aut}(\mathsf{B}_e) $ is an irreducible permutation representation.
 \end{enumerate}
 We expect an irreducible $\alpha^{-1}$-twisted $G$-crossed extension is precisely a Lagrangian algebra $L_\lambda \in \mathsf{Z}(\C)$ corresponding to the Neumann boundary condition $\cN_\lambda$.\footnote{Lagrangian algebras in braided fusion 2-categories were defined in~\cite{decoppet2024local}.}


\subsection{Formulation as 3-representations}

We now propose that $\alpha^{-1}$-twisted $G$-crossed extensions are in 1-1 correspondence with $\alpha^{-1}$-projective 3-representations of $G$. There is much additional structure compared to the analogous construction in two dimensions because braided automorphisms of $\mathsf{B}_e$ include not only permutations of the summands but act by internal braided automorphisms of each summand $\mathsf{b}_j$.

\subsubsection{General 3-representations}

We view an $\alpha^{-1}$-projective 3-representation as a 3-group homomorphism 
\be
\lambda : \mathsf{G}_\alpha \longrightarrow \mathsf{Aut}(\mathsf{B}_e) \, ,
\ee
where:
\begin{itemize}
	\item The source is the 3-group extension with homotopy groups
	\bea
	\pi_0(\mathsf{G}_\alpha)  & = G  \\
	\pi_1(\mathsf{G}_\alpha)  & = 1   \\
	\pi_2(\mathsf{G}_\alpha)  & = \bC^\times
	\eea
	and Postnikov invariant $[\alpha^{-1}]\in H^4(G, \bC^\times)$.
	\item The target is the 3-group of braided automorphisms that captures invertible symmetries and 't Hooft anomalies of the 3-dimensional oriented TQFT. It has homotopy groups
	\bea
	\pi_0(\mathsf{Aut}(\mathsf{B}_e))  & = \text{Aut}(\mathsf{B}_e)  \\
	\pi_1(\mathsf{Aut}\mathsf{B}_e)) & = (\mathsf{B}_e)^\times \\
	\pi_2(\mathsf{Aut}(\mathsf{B}_e))  & = (\bC^\times)^{|\B_e|}
	\eea
	where $(\B_e)^\times$ is the Picard group of invertible lines and $|\B_e| := n$ is the number of summands in the decomposition~\eqref{eq:braided-decomposition}.
	In addition, there is Postnikov data
	\bea
	\,[P_3] & \in H^3(\text{Aut}(\mathsf{B}_e), (\mathsf{B}_e)^\times) \, , \\
	\,[P_4] & \in H^4(\pi_{\leq 1},U(1)^{|\B_e|}) \, ,
	\eea
	where $\pi_{\leq 1}$ is shorthand for the 2-group truncation with homotopy groups $\text{Aut}(\mathsf{B}_e), (\mathsf{B}_e)^\times$ and Postnikov class $[P_3]$.
\end{itemize}
The potentially non-trivial Postnikov data is a new ingredient compared to two dimensions and will feature prominently in what follows.

An $\alpha^{-1}$-projective 3-representation on $\mathsf{B}_e$ is now determined by
\begin{enumerate}
\item A homomorphism $\Sigma : G \to \text{Aut}(\B_e)$.
\item A 2-cochain $\psi \in C^2(G,(\mathsf{B}_e)^\times)$ satisfying $\delta\psi = O_3$.
\item A 3-cochain $c \in C^3(G,U(1)^{|\B_e|} )$  satisfying $\delta c = \alpha^{-1} O_4 $.
\end{enumerate}
where
\bea
O_3 & := \Sigma^* P_3 \, , \\
O_4 & := (\Sigma,\psi)^* P_4 \, .
\eea
denote obstructions to the $G$-action on $\mathsf{B}_e$. In forming the second obstruction $O_4$, we view the combination of $\Sigma, \psi$ as a group homomorphism $(\Sigma,\psi): G \to \pi_{\leq 1}$. Before unpacking the definition further, we now simplify to irreducible 3-representations.

\subsubsection{Irreducible 3-representations}

For an irreducible 3-representation, we must demand
\be
\mathsf{B}_e \cong \mathsf{b}^n := \mathsf{b} \oplus \cdots \oplus \mathsf{b} \, ,
\ee
such that
\be
\mathsf{Aut}(\mathsf{B}_e) = \mathsf{Aut}(\mathsf{b})^n \rtimes S_n 
\ee
where $S_n$ acts by permutations on $\mathsf{Aut}(\mathsf{b})^n$. We can now proceed in steps by first considering automorphisms of each summand and then combining with permutations.

First, $\mathsf{Aut}(\mathsf{b})$ captures the invertible symmetries and anomalies of the irreducible theory determined by $\mathsf{b}$. It has homotopy groups
\bea
\pi_0(\mathsf{Aut}(\mathsf{b})) & = \text{Aut}(\mathsf{b})  \\
\pi_1(\mathsf{Aut}(\mathsf{b}))  & = \mathsf{b}^\times \\
\pi_2(\mathsf{Aut}(\mathsf{b}))  & = \bC^\times \, ,
\eea
corresponding to an invertible 0-form symmetry $\text{Aut}(\mathsf{b})$ and 1-form symmetry $\mathsf{b}^\times$. The Postnikov data
\bea
\, [p_3] & \in H^3(\text{Aut}(\mathsf{b}), \mathsf{b}^\times) \, , \\
\, [p_4] & \in H^3(\pi_{\leq 1}, U(1) ) 
\eea
captures the 2-group structure and 't Hooft anomaly respectively. 

Second, we combine with permutations to find 
\bea
\pi_1(\mathsf{Aut}(\mathsf{B}_e) ) & = \text{Aut}(\mathsf{b})^n \rtimes S_n \\
\pi_2(\mathsf{Aut}(\mathsf{B}_e) ) & = (\mathsf{b}^\times)^n \\
\pi_3(\mathsf{Aut}(\mathsf{B}_e) ) & = (\bC^\times)^n
\eea
where $S_n$ acts by permutations on $\text{Aut}^{br}(\mathsf{b})^n,(\mathsf{b}^\times)^n,U(1)^n$. The combination has Postnikov data $[P_3]$, $[P_4]$ determined by $[p_3]$, $[p_4]$.

With this decomposition in hand, the data of an irreducible $\alpha^{-1}$-projective 3-representation on $\mathsf{B}_e = \mathsf{b}^n$ can be written down explicitly as:
\begin{enumerate}
\item A homomorphism $\Sigma : G \to \text{Aut}(\mathsf{b})^n \rtimes S_n$ is a pair $(\rho^j,\sigma)$
satisfying
\be
(\rho^{\sigma_h(j)}_g,\sigma_g) \circ (\rho^j_g,\sigma_h) = (\rho^j_{gh},\sigma_{gh}) 
\ee
where $\sigma : G \to S_n$ is an irreducible permutation representation.
\item A collection $\psi_j : G \times G \to \mathsf{b}^\times$ satisfying
\be
{\rho^{\sigma_g^{-1}(j)}_g( \psi_{\sigma_g^{-1}(j)}(h,k))\psi_j(g,hk)} = {\psi_j(gh,k)\psi_j(g,h)} O_3 (g,h,k) \, .
\ee
\item A collection $c_j : G \times G \times G \to U(1)$ satisfying
\be
c_{\sigma_g^{-1}(j)}(h,k,l)c_j(g,hk,l)c_j(g,h,k) = c_j(gh,k,l)c_j(g, h, kl)\frac{O_4(g,h,k,l)}{\alpha(g, h, k,l)} \, .
\ee
\end{enumerate}
These equations must be satisfied for all $j=1,\ldots,n$ and $g,h,k,l \in G$.

\subsubsection{Constructing the Extension}

A $\alpha^{-1}$-twisted $G$-crossed extension can constructed from an $\alpha^{-1}$-projective 3-representation of $G$ as follows.
The graded components are
\be
\mathsf{B}_g = \sum_{j \, | \, \sigma_g(j) = j} \mathsf{b} \cdot e^g_j
\ee
with generators $e_g^j$ with $\sigma_g(j) = j$ such that
\be
\rho_{g}(h) :  \mathsf{b} \cdot e^j_h  \longrightarrow   \rho_g^j( \mathsf{b} ) \cdot e^{\sigma_g(j)}_{ghg^{-1}} 
\ee
with compositor data $\rho^\circ$ and distributor data $\rho^\otimes$ determined by $\psi$, and associator data determined by $c$. All such irreducible $\alpha^{-1}$-projective 3-representations or $\alpha^{-1}$-twisted $G$-crossed extensions can be obtained by induction.


\subsection{Subgroups and Induction}

We now describe an equivalent presentation of this data using induction of 3-representations from subgroups. The associated gauging operation corresponds to stacking with an irreducible 3-dimensional oriented TQFT equipped with symmetry structure $(H,(\alpha|_H)^{-1})$ and then gauging $H \subseteq G$. 

The starting point is:
\begin{enumerate}
\item A subgroup $H\subseteq G$.
\item A modular fusion category $\mathsf{b}$.
\item An $(\alpha|_H)^{-1}$-twisted $H$-crossed extension of $\mathsf{b}$.
\end{enumerate}
We emphasise that the final data point is a projective 3-representation on a modular fusion category $\mathsf{b}$ ($n=1$) that we will induce from $H$ to $G$. 

Unpacking this data gives:
\begin{enumerate}
\item A subgroup $H \subseteq G$.
\item A modular fusion category $\mathsf{b}$.
\item A homomorphism $\rho : H \to \text{Aut}(\mathsf{b})$.
\item A 2-cochain $\psi \in C^2(H,\mathsf{b}^\times)$ satisfying $\delta\psi = o_3$.
\item A 3-cochain $\phi \in C^3(H,U(1))$ satisfying $ \delta \phi = (\alpha|_H)^{-1} o_4 $.
\end{enumerate}
where $[o_3]$, $[o_4]$ are the obstructions determined sequentially as above.

This data determines an irreducible $\alpha^{-1}$-projective 3-representation of $G$ by induction. 
To construct it explicitly, we again choose representatives of left cosets $a_j H$ to determine a permutation representation on $G/H \cong \{a_1,\ldots,a_n\}$ by 
$g \cdot a_j H = a_{\sigma_g(j)} H$ with compensating transformations $\ell_{g,j} = a_{\sigma_g(j)}^{-1} \cdot g \cdot a_j$.

The induced projective 3-representation on $\mathsf{b}^n$ is:
\begin{itemize}
\item An irreducible representation $\Sigma = (\rho^j,\sigma)$ determined by
\be
\rho^j_g = \rho(\ell_{g, \sigma^{-1}_g(j)})\, .
\ee
\item A 2-cochain $\psi \in C^2(G,(\mathsf{b}^\times)^n)$ with components
\be
\psi_j(g_1,g_2) \propto \psi(\ell_{g_1,\sigma^{-1}_{g_1}(j)},\ell_{g_2,\sigma^{-1}_{g_1g_2}(j)}) \, .
\label{eq:3d-induced-psi}
\ee
\item A 3-cochain $c \in C^3(G,U(1)^n)$ with components
\be
c_j(g_1,g_2,g_3) \propto \phi(\ell_{g_1,\sigma^{-1}_{g_1}(j)},\ell_{g_2,\sigma^{-1}_{g_1g_2}(j)},\ell_{g_3, \sigma^{-1}_{g_1g_2g_3}(j)}) \, .
\label{eq:3d-induced-c}
\ee
\end{itemize}
In writing equations~\eqref{eq:3d-induced-psi} and~\eqref{eq:3d-induced-c} we have again omitted factors that are independent of $\psi$ and $\phi$ respectively. 

The associated $\alpha^{-1}$-twisted $G$-crossed extension are similarly determined by induction and admits a decomposition 
with graded components
\be
\mathsf{B}_g  =  \bigoplus_{j \, | \, g \in {}^{a_j}H} \mathsf{b} \cdot e_g^j \, .
\ee
with generators $e^j_g$ with $g \in {}^{a_j}H$ such that
\be
\rho_g(h) : \mathsf{b} \cdot e_h^j \to  \rho_g^j( \mathsf{b} ) \cdot  e^{\sigma_g(j)}_{ghg^{-1}} 
\ee
with compositor data $\rho^\circ$ and distributor data $\rho^\otimes$ again determined by $\psi$, and associator data determined by $c$. All such irreducible $\alpha^{-1}$-projective 3-representations or $\alpha^{-1}$-twisted $G$-crossed extensions can be obtained by induction.

The associated gauging operation is now equivalent to stacking with an irreducible 3-dimensional oriented TQFT equipped with symmetry structure $(H,(\alpha|_H)^{-1})$ and then gauging $H \subseteq G$. Following our notation in two dimensions, we can represent this by
\be
\cT  \longrightarrow \cT /\!_\Psi \, H  \, ,
\ee
where $\Psi$ denotes the collection of data $(\mathsf{b},\rho, \psi, \phi)$. This gauging operation shifts
\be
\zeta_{\cT /\!_\Psi \, H} =   \zeta_\cT \cdot \xi(\mathsf{b})
\ee
where $\xi(\mathsf{b})$ is the multiplicative central charge of $\mathsf{b}$. There is no physical distinction between conjugate subgroups $H$, ${}^aH$, which induce equivalent 3-representations. 

All $\alpha^{-1}$-projective 3-representations of $G$ or $\alpha^{-1}$-twisted $G$-crossed extensions can be obtained in this way by induction and so we propose this provides a classification of gauging operations, gapped systems, or gapped boundary conditions.

\subsubsection{Example 1}

The simplest class of examples correspond to demanding $\mathsf{b} = \mathsf{Vect}$ such that there are no automorphisms and obstructions vanish. The remaining data reduces to
\begin{enumerate}
\item A subgroup $H \subseteq G$.
\item A 3-cocycle $\phi \in C^3(H,\bC^\times)$ satisfying $ \delta \phi = (\alpha|_H)^{-1}$.
\end{enumerate}
This requires $H \subseteq G$ is anomaly free and $\phi$ is a local counter-term trivialising the anomaly.
This is the most direct lift of the construction in two dimensions and reproduces the class of gauging operations considered in~\cite{Bartsch:2022mpm,Bartsch:2022ytj,Bhardwaj:2022lsg,Bhardwaj:2022maz}. 

\subsubsection{Example 2}

A broader class of examples correspond to choosing an anomaly free subgroup $H\subseteq G$ with $[\alpha|_H] = 0$ and a $H$-crossed extension $\mathsf{B}$ of a modular fusion category $\mathsf{b}$.

Choosing trivialisations of the obstructions $o_3$, $o_4$ and shifting $\psi, \phi$ appropriately the data can be summarised as follows
\begin{enumerate}
\item A subgroup $H \subseteq G$.
\item A modular fusion category $\mathsf{b}$.
\item An homomorphism $\rho : H \to \text{Aut}(\mathsf{b})$.
\item A 2-cocycle $\psi \in C^2(H,\mathsf{b}^\times)$ satisfying $\delta\psi = 0$.
\item A 3-cocycle  $\phi \in C^3(H,\bC^\times)$ satisfying $ \delta \phi = (\alpha|_H)^{-1}$.
\end{enumerate}
From the perspective of gapped systems, this is spontaneous symmetry breaking combined with the classification of~\cite{Barkeshli_2019}. 

Furthermore, specialising to $\alpha = 0$, Lagrangian algebras $L \in \mathsf{Z}(\C)$ are known to be classified by a subgroup $H\subseteq G$ and a modular extension of $\mathsf{Rep}(H)$~\cite{decoppet2024local}. This is compatible since a $H$-crossed extension determines a modular extension
\be
\mathsf{Rep}(H) \hookrightarrow \mathsf{B}^H \, ,
\ee
where $\mathsf{B}^H$ denotes the $H$-equivariantization of $\mathsf{B}$, and conversely. This reproduces the gauged perspective on the classification of gapped systems with symmetry in 2+1 dimensions developed in~\cite{Lan:2016rcq,Lan:2017krb}.


\subsection{Symmetry Categories}

Let us now consider the symmetry category of the gauged theory $\cT /\!_\psi\, H$ or Neumann boundary condition $\cN_\lambda$. This should be determined by computing module categories over the associated Lagrangian algebra,
\be
\mathsf{C}_{\lambda} := \mathsf{Mod}_{\mathsf{Z}(\C)}(L_\lambda) \, .
\ee
A general analysis of the structure of this spherical fusion 2-category is beyond the scope of this paper and we limit ourselves to some basic observations for the two classes of examples considered above.

\subsubsection{Example 1}

For examples indexed by an anomaly-free subgroup $H \subseteq G$ and trivialisation $ \delta \phi = (\alpha|_H)^{-1}$, the symmetry category is a group-theoretical fusion 2-category
\be
\mathsf{C}_{\lambda} = \mathsf{C}(G,\alpha | H , \phi ) \, ,
\ee
whose structure was studied in~\cite{Bartsch:2022mpm,Bartsch:2022ytj,Bhardwaj:2022lsg,Bhardwaj:2022maz}.

There is an inclusion
\be
\mathsf{2Rep}(H) \hookrightarrow \mathsf{C}_{\lambda}
\ee
generated by topological Wilson lines and their condensations. In addition, it admits a decomposition as a 2-category
\be
\mathsf{C}_{\lambda} \cong \bigoplus_{[g] \in H\backslash G /H} \mathsf{2Rep}^{c_g}(H \cap {}^g H) \, ,
\label{eq:3d-dec1}
\ee
where the summation is over double cosets with representatives $g \in G$ and $c_g \in Z^3(H \cap {}^g H,\mathbb{C}^\times)$ is a certain 3-cocycle built from $\alpha$, $\psi$. This provides an enumeration of simple objects. The fusion rules are then built on a foundation of the double coset ring, supplemented with rules for decomposing and combining projective 2-representations. 

\subsubsection{Example 2}

Now consider the broader class of examples indexed by an anomaly-free subgroup $H \subseteq G$ and a $H$-crossed extension $\mathsf{B}$ of a modular fusion category $\mathsf{b}$. 

There is now an inclusion
\be
\mathsf{Mod}(\mathsf{b}^H) \hookrightarrow \mathsf{C}_{\lambda}
\ee
where the equivariantization $\mathsf{b}^H$ satisfies $\mathsf{Z}_{(2)}(\mathsf{b}^H) = \mathsf{Rep}(H)$.
This clearly reduces to the previous example when $\mathsf{b} = \mathsf{Vect}$. 

Furthermore, a preliminary examination suggests a decomposition as a 2-category
\be
\mathsf{C}_{\lambda} \cong \bigoplus_{[g] \in H\backslash G /H} \mathsf{Mod}(\mathsf{B}|_{H \cap {}^g H}) \, ,
\ee
where $\mathsf{B}|_{H \cap {}^g H}$ denotes the fusion category obtained by restricting the $H$-crossed extension to $H \cap {}^gH$. As a consistency check, when $\mathsf{b} = \mathsf{Vect}$, 
\be
\mathsf{B}|_{H \cap {}^g H} = \mathsf{Vect}^{c_g}(H \cap {}^g H) \, ,
\ee
and this reproduces the previous example.
This should provide a way to enumerate simple objects and form a foundation for expressing the fusion structure.



\acknowledgments

MB is supported by the EPSRC Early Career Fellowship EP/T004746/1 ``Supersymmetric Gauge Theory and Enumerative Geometry", STFC Research Grant ST/T000708/1 ``Particles, Fields and Spacetime" and the Simons Collaboration on Global Categorical Symmetry. 



\bibliographystyle{JHEP}
\bibliography{references}
\end{document}

%% file: preamble.tex
\usepackage{amsfonts}
\usepackage{amscd}
\usepackage{amssymb}
\usepackage{amsmath,bbm}
\usepackage{graphicx}
\usepackage{epsfig}
\usepackage{latexsym}
\usepackage{mathtools}
\usepackage{hyperref}
\usepackage{tikz-cd}
\usepackage[vcentermath]{youngtab}

\def \be  {\begin{equation}}
\def \ee  {\end{equation}}
\def \bea {\begin{equation}\begin{aligned}}
\def \eea {\end{aligned}\end{equation}}
\def \ba  {\begin{eqnarray}}
\def \ea  {\end{eqnarray}}
\def \bb  {\begin{thebibliography}}
\def \eb  {\end{thebibliography}}
\def \lab #1 {\label{#1}}

\newcommand\cD{\mathcal{D}}

\newcommand\cN{\mathcal{N}}

\newcommand\cT{\mathcal{T}}

\newcommand\C{\mathsf{C}}
\newcommand\B{\mathsf{B}}

\newcommand\bC{\mathbb{C }}

\definecolor{cardinal}{rgb}{0.6,0,0}
\definecolor{darkgreen}{rgb}{0,0.5,0}
\definecolor{golden}{rgb}{0.92, 0.7, 0}
\definecolor{midnight}{rgb}{0, 0, 0.5}
\definecolor{darkblue}{rgb}{0.2, 0, 0.8}